\documentclass[12pt,a4paper]{article}
\usepackage{amsmath,amssymb,graphics,multirow,multicol,anysize,rotating,setspace}
\numberwithin{equation}{section}
\title{\textbf{ Power Maxwell distribution: Statistical Properties, Estimation and Application}}
\author{{\small Abhimanyu Singh Yadav$^{1}$\footnote{Corresponding author E-mail: asybhu10@gmail.com}, Hassan S. Bakouch $^2$, Sanjay Kumar Singh$^3$ and Umesh Singh$^3$}\vspace{0.5cm}\\$^1${\small Department of Statistics, Central University of Rajasthan, Ajamer, India\vspace{0.2cm}}\\$^2$ {\small Department of Mathematics, Faculty of Science, Tanta University, Tanta, Egypt\vspace{0.2cm}}\\$^3${\small Department of Statistics and DST-CIMS, BHU, Varanasi, India}}
\date{}
\begin{document}
	\maketitle
	\begin{abstract}	
In this article, we proposed a new probability distribution named as power Maxwell distribution (PMaD). It is another extension of Maxwell distribution (MaD) which would lead more flexibility to analyze the data with non-monotone failure rate. Different statistical properties such as reliability characteristics,  moments, quantiles, mean deviation, generating function, conditional moments, stochastic ordering, residual lifetime function and various entropy measures have been derived. The estimation of the parameters for the proposed probability distribution has been addressed by maximum likelihood estimation method and Bayes estimation method. The Bayes estimates are obtained under gamma prior using squared error loss function. Lastly, real-life application for the proposed distribution has been illustrated through different lifetime data. 
	\end{abstract}
\section{Introduction}
The Maxwell distribution has broad application in statistical physics, physical chemistry, and their related areas. Besides Physics and Chemistry it has good number of applications in reliability theory also. At first, the Maxwell distribution was used as lifetime distribution by Tyagi and Bhattacharya (1989). The inferences based on generalized Maxwell distribution has been discussed by Chaturvedi and Rani (1998).  Bekker and Roux (2005) consider the estimation of reliability characteristics under for Maxwell distribution under Bayes paradigm. Radha and Vekatesan (2005) discuss the prior selection procedure in case of Maxwell probability distribution. Shakil et al. (2008) studied the distributions of the product $|XY|$ and ratio $|X/Y|$ when X and Y are independent random variables having the Maxwell and Rayleigh distributions. Day and Maiti (2010) proposed the Bayesian estimation of the parameter for the Maxwell distribution. Tomer and Panwar (2015) discussed the estimation procedure for the parameter of Maxwell distribution in the presence of progressive type-I hybrid censored data. After this, Modi (2015), Saghir and Khadim (2016), proposed lengths biased Maxwell distribution and discussed its various properties. Furthermore, several generalizations based on Maxwell distribution are advocated and statistically justified. Recently, two more extensions of Maxwell distribution has been introduced by Sharma et al. (2017a), (2017b) and discussed the classical as well as Bayesian estimation of the parameter along with the real-life application.\\

A random variable $Z$ follows  Maxwell distribution with scale parameter $\alpha$, denoted as $Z \sim
MaD(\alpha )$, if its probability density function (pdf) and cumulative distribution function (cdf) are
given by
\begin{equation}
f(z,\alpha)=\dfrac{4}{\sqrt{\pi}} \alpha^{\frac{3}{2}} z^{2}e^{-\alpha z^{2}}~~~~z\ge 0,\alpha>0
\end{equation}
and
\begin{equation}
F(z,\alpha)=\dfrac{2}{\sqrt{\pi}}\Gamma\left( \frac{3}{2},\alpha z^{2}\right) 
\end{equation}
respectively, where $\Gamma (a, z) =\int_{0}^{z}p^{a-1}e^{-p}dp$  is the incomplete gamma function.\\

In this article, we proposed PMaD as a new generalization of the Maxwell distribution and discussed its various statistical properties and application. The objective of this article is to study the statistical properties of the PMaD distribution, and then estimate the unknown parameters using classical and Bayes estimation methods. Other motivations regarding advantages of the PMaD distribution comes from its flexibility to model the data with nono-monotone failure rate. Thus, it can be taken as an excellent alternative to several inverted families of distributions. The uniqueness of this study comes from the fact that we provide a comprehensive description of mathematical and statistical properties of this distribution with the hope that it will attract more extensive applications in biology, medicine, economics, reliability, engineering, and other areas of research.\\

The rest of the paper has been shaped in the following manner. The introduction of the proposed study including the methodological details is given in Section and Subsection of 1.  Section 2 provides some statistical properties related to the proposed model. Residual and reverse residual lifetime function for PMaD is derived in Section 3. In Section 4, order statistics have been obtained. The MLEs and Bayes estimation procedure have been discussed in Section 5. In Section 6,  simulation study is carried out to compare the performance of maximum likelihood estimates (MLEs) and Bayes estimates. In Section 7, we illustrate the application and usefulness of the proposed model by applying it to four data sets. Finally, Section 8 offers some concluding remarks.

\section{Power Maxwell Distribution and Statistical Properties}
In statistical literature, several generalizations based on certain baseline probability distribution have been advocated regarding the need of the study. These generalized model accommodate the various nature of hazard rate and seems to be more flexible. Here, this paper provides another generalization of the MaD using power transformation of Maxwell random variates. Let us consider a transformation $X=Z^{\frac{1}{\beta}}$ where $Z\sim MaD(\alpha)$. Then the resulting distribution of $X$ is called as  power maxwell distribution with parameter $\alpha$ and $\beta$ respectively. From now, it is denoted by $X\sim PMaD(\alpha,\beta)$, where, $\alpha$ and $\beta$ are the scale and shape parameter of the proposed distribution. The probability density function and cumulative distribution function of the PMaD are given by 
\begin{equation}
f(x,\alpha,\beta)=\dfrac{4}{\sqrt{\pi}} \alpha^{\frac{3}{2}}\beta x^{3\beta-1}e^{-\alpha x^{2\beta}}~~~~x\ge 0,\alpha,\beta>0
\end{equation}
\begin{equation}
F(x,\alpha,\beta)=\dfrac{2}{\sqrt{\pi}}\gamma\left( \frac{3}{2},\alpha x^{2\beta}\right) 
\end{equation}
respectively. The different mathematical and statistical properties such as moments, reliability, hazard, median, mode, the coefficient of variation, mean deviation, conditional moments, Lorentz curve, stochastic ordering, residual life, entropy measurements,  of PMaD have been derived in following subsections. 
\begin{figure}
	\includegraphics[width=6.5in,height=4in]{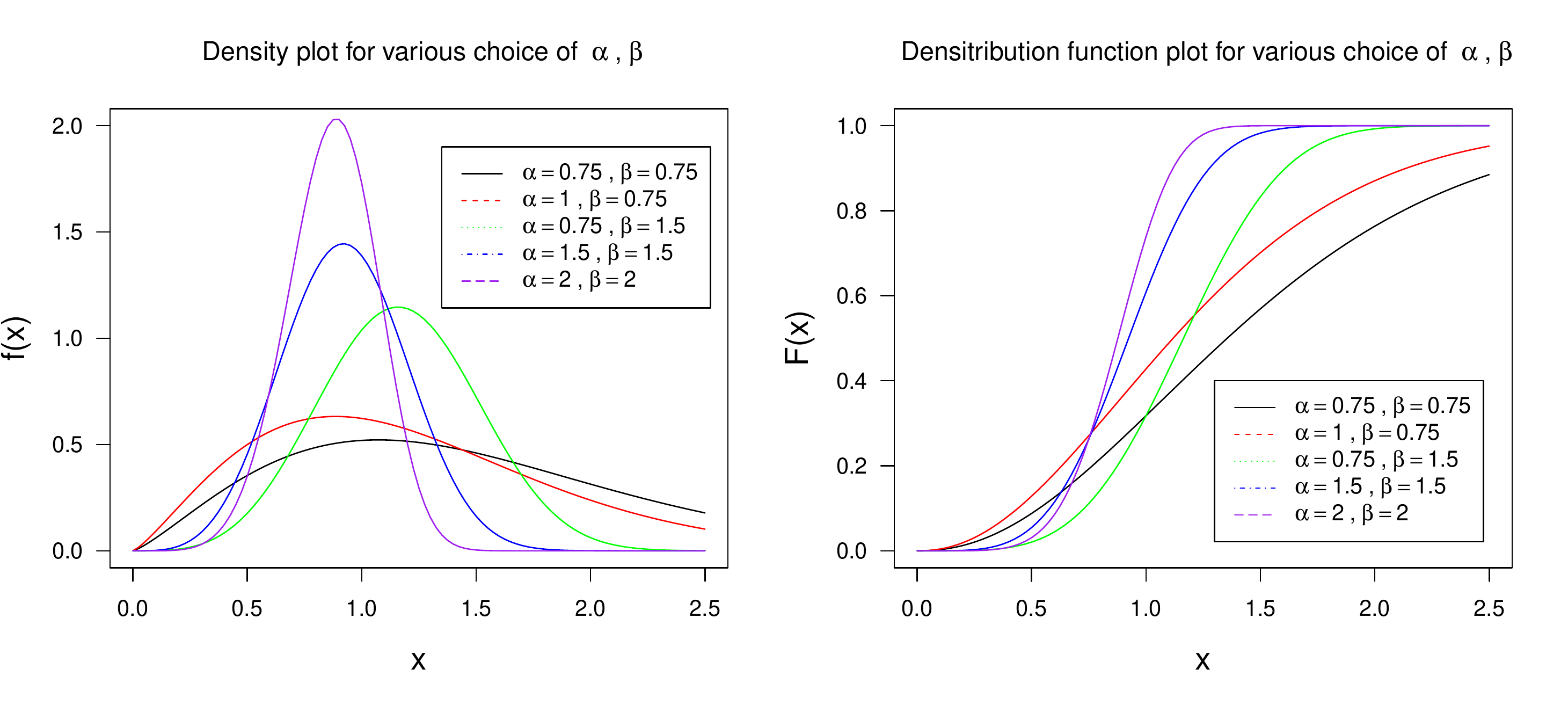}
	\end{figure}
\subsection{Asymptotic behaviour}
This subsection, described the symptotic nature of density and survival functions for the proposed distribution. To illustrate assymptoic behaviour, at first, we will show that $\lim\limits_{x\rightarrow 0}f(x,\alpha,\beta)=0$ and $\lim\limits_{x\rightarrow \infty}f(x,\alpha,\beta)=0$ . Therefore, using (2.1) 
\begin{equation*}
\begin{split}
\lim\limits_{x\rightarrow 0}f(x,\alpha,\beta)&=\dfrac{4}{\sqrt{\pi}} \alpha^{\frac{3}{2}}\beta\lim\limits_{x\rightarrow 0} x^{3\beta-1}e^{-\alpha x^{2\beta}}\\&=\dfrac{4}{\sqrt{\pi}} \alpha^{\frac{3}{2}}\beta\times 0=0
\end{split}
\end{equation*}
$\implies$ $\lim\limits_{x\rightarrow \infty}f(x,\alpha,\beta)=0$\\

and 
\begin{equation*}
\begin{split}
\lim\limits_{x\rightarrow \infty}f(x,\alpha,\beta)&=\dfrac{4}{\sqrt{\pi}} \alpha^{\frac{3}{2}}\beta\lim\limits_{x\rightarrow \infty} x^{3\beta-1}\lim\limits_{x\rightarrow \infty}e^{-\alpha x^{2\beta}}\\&=\dfrac{4}{\sqrt{\pi}} \alpha^{\frac{3}{2}}\beta\times \infty\times 0=0
\end{split}
\end{equation*}
$\implies$ $\lim\limits_{x\rightarrow \infty}f(x,\alpha,\beta)=0$\\

Similarly, the asymptotic behaviour of survival function can also be shown and found  that $\lim\limits_{x\rightarrow 0}S(x)=1$ and  $\lim\limits_{x\rightarrow \infty}S(x)=0$.  
\subsection{Reliability and hazard functions}
The characteristics based on reliability function and hazard function are very useful to study the pattern of any lifetime phenomenon. The reliability and hazard function of the proposed distribution have been derived as;
\begin{itemize}
	\item The reliability function $R(x,\alpha,\beta)$ is given by
	\begin{equation}
	R(x)=1-\dfrac{2}{\sqrt{\pi}}\gamma\left( \frac{3}{2},\alpha x^{2\beta}\right) 
	\end{equation}
	\item  The mean time to system failure (MTSF) is given as
		\begin{equation}
	Mt(x)=\dfrac{2}{\sqrt{\pi}}\left( \frac{1}{\alpha}\right) ^{\frac{1}{2\beta}}\Gamma\left( \dfrac{3\beta+1}{2\beta}\right)  
	\end{equation}
	\item  The hazard function $H(x)$ is given as
		\begin{equation}
	H(x)=\dfrac{4\alpha^{\frac{3}{2}}\beta x^{3\beta-1}e^{-\alpha x^{2\beta}}}{\sqrt{\pi}-2\gamma\left( \frac{3}{2},\alpha x^{2\beta}\right) }
	\end{equation}
	\begin{figure}
		\includegraphics[width=6.5in,height=4in]{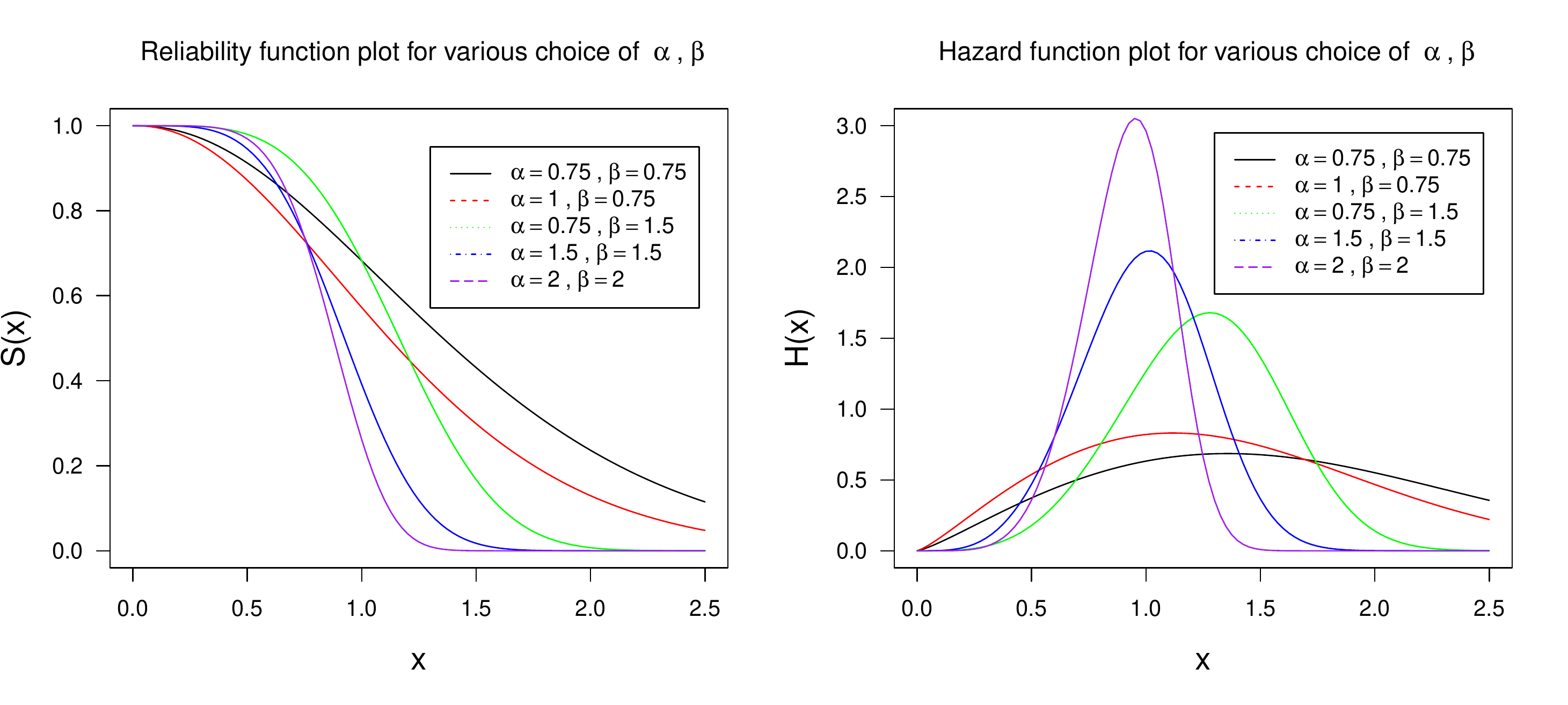}
	\end{figure}
	\item The reverse hazard rate $h(x)$ is obatined as
	\begin{equation}
      h(x)=\dfrac{2\alpha^{\frac{3}{2}}\beta x^{3\beta-1}e^{-\alpha x^{2\beta}}}{\gamma\left( \frac{3}{2},\alpha x^{2\beta}\right) }
\end{equation}
     \item The odds function is defined as;
     \begin{equation}
     O(x)=\dfrac{2\gamma\left( \frac{3}{2},\alpha x^{2\beta}\right) }{\sqrt{\pi}-2\gamma\left( \frac{3}{2},\alpha x^{2\beta}\right) }
     \end{equation}
\end{itemize} 
\subsection{Moments}
Let $x_1,x_2,\cdots x_n$ be the random observation from PMaD$(\alpha,\beta)$. The $r^{th}$ moment $\mu_r^{'}$ about origin is defined as 
\begin{equation}
\begin{split}
\mu_r^{'}&=E(x^{r})=\int_{x=0}^{\infty}x^{r}f(x,\alpha,\beta)\quad dx\\&=\dfrac{2}{\sqrt{\pi}}\left( \dfrac{1}{\alpha}\right)^{\frac{r}{2\beta}}\Gamma\left( \dfrac{3\beta+r}{2\beta}\right)  
\end{split}
\end{equation} 
The first, second, third and fourth raw moment about origin are obtained by putting $r=1, 2 ,\cdots,4$ in above expression. If $r=1$ then we get mean of the proposed distribution. Thus,
\begin{equation}
\mu_1^{'}=\dfrac{2}{\sqrt{\pi}}\left( \frac{1}{\alpha}\right) ^{\frac{1}{2\beta}}\Gamma\left( \dfrac{3\beta+1}{2\beta}\right)  
\end{equation}
for $r=2,3 \, \& 4$
\begin{equation}
\mu_2^{'}=\dfrac{2}{\sqrt{\pi}}\left( \frac{1}{\alpha}\right) ^{\frac{1}{\beta}}\Gamma\left( \dfrac{3\beta+2}{2\beta}\right)  
\end{equation}
\begin{equation}
\mu_3^{'}=\dfrac{2}{\sqrt{\pi}}\left( \frac{1}{\alpha}\right) ^{\frac{3}{2\beta}}\Gamma\left( \dfrac{3\beta+3}{2\beta}\right) 
\end{equation}
and 
\begin{equation}
\mu_4^{'}=\dfrac{2}{\sqrt{\pi}}\left( \frac{1}{\alpha}\right) ^{\frac{2}{\beta}}\Gamma\left( \dfrac{3\beta+4}{2\beta}\right)  
\end{equation} 
The respective central moment can be evaluated by using the following relations.
\begin{equation}
\mu_2=\mu_2^{'}-\left( \mu_1^{'}\right)^{2}=\dfrac{4}{\pi}\left( \dfrac{1}{\alpha}\right)^{\frac{1}{\beta}}\left[ \sqrt{\frac{\pi}{4}} \Gamma\left( \dfrac{3\beta+2}{2\beta}\right) -\left\lbrace \Gamma\left( \dfrac{3\beta+1}{2\beta}\right)\right\rbrace^{2} \right] 
\end{equation}
\begin{equation}
\mu_3=\mu_3^{'}-3\mu_2^{'}\mu_1^{'}+2\left( \mu_1^{'}\right) ^{3}
\end{equation}
\begin{equation}
\mu_4=\mu_4^{'}-4\mu_3^{'}\mu_1^{'}+6\mu_2^{'}\left( \mu_1^{'}\right) ^{2}-3\left( \mu_1^{'}\right) ^{4}
\end{equation}
\subsection{Coefficient of Skewness and Kurtosis}
The coefficient of skewness and kurtosis  measure nd convexity of the curve and its shape. It is obtained by moments based relations suggested by Pearson and given by;
\begin{equation}
\beta_1=\dfrac{\left[ \mu_3^{'}-3\mu_2^{'}\mu_1^{'}+2\left( \mu_1^{'}\right) ^{3}\right] ^{2}}{\left[\mu_2^{'}-\left( \mu_1^{'}\right)^{2} \right]^{3} }
\end{equation}
and
\begin{equation}
\beta_2=\dfrac{\mu_4^{'}-4\mu_3^{'}\mu_1^{'}+6\mu_2^{'}\left( \mu_1^{'}\right) ^{2}-3\left( \mu_1^{'}\right) ^{4}}{\left[\mu_2^{'}-\left( \mu_1^{'}\right)^{2} \right]^{2} }
\end{equation}
These values are calculated in Table 1 for different combination of  model parameters and it is observed that the shape of PMaD is right skewed and almost symmetrical for some choices of $\alpha, \beta$. Also, it can has the nature of platykurtic, mesokurtic and leptokurtic, thus PMaD may be used to model skewed and symmetric data as well. 
\subsection{Coefficient of variation}
The coefficient of variation (CV) is calculated by 
\begin{equation}
CV=\dfrac{\sqrt{\mu_2^{'}-\left( \mu_1^{'}\right)^{2}}}{\mu_1^{'}}\times 100
\end{equation}
\subsection{Mode and Median}
The mode $(M_0)$ for PMaD $(\alpha,\beta)$ is obtained by solving the following expression
\begin{equation}
\dfrac{d}{dx}f(x,\alpha,\beta)|_{M_0}=0
\end{equation}
which yield 
$$M_0=\left( \dfrac{3\beta-1}{2\alpha\beta}\right)^{\frac{1}{2\beta}} $$
The median $(M_{d})$ of the proposed distribtuion can be calculated by using the empirical relation amung mean, median and mode. Thus, the median is,
$$M_{d}=\frac{1}{3} M_{0}+\frac{2}{3}\mu_1^{'}=\frac{1}{3} \left[ \left( \dfrac{3\beta-1}{2\alpha\beta}\right)^{\frac{1}{2\beta}}+ \dfrac{4}{\sqrt{\pi}}\left( \frac{1}{\alpha}\right) ^{\frac{1}{2\beta}}\Gamma\left( \dfrac{3\beta+1}{2\beta}\right)\right]   $$
\begin{table}[htbp]
	\centering
	\caption{Values of mean, variance, skewness, kurtosis, mode and coefficient of variation  for different $\alpha$, $\beta$}
	\begin{tabular}{ccccccc}\\
		\hline
		\multirow{2}[0]{*}{$ \alpha,~ \beta $} & $\mu_1^{'}$ & $\mu_2$ & $\beta_1$& $\beta_2$ & $x_0$  & CV \\
		\cline{2-7}		
		& \multicolumn{6}{c}{when alpha fixed and beta varying} \\
		\hline
		0.5, 0.5 & 3.0008 & 5.9992 & 2.6675 & 7.0010 & 1.0000 & 0.8162 \\
		0.5, 1.0 & 1.5962 & 0.4530 & 0.2384 & 3.1071 & 1.4142 & 0.4217 \\
		0.5, 1.5 & 1.3376 & 0.1499 & 0.0102 & 2.7882 & 1.3264 & 0.2894 \\
		0.5, 2.5 & 1.1780 & 0.0445 & 0.0481 & 2.7890 & 1.2106 & 0.1792 \\
		0.5, 3.5 & 1.1204 & 0.0211 & 0.1037 & 2.4351 & 1.1533 & 0.1298 \\
		\hline
		\multicolumn{7}{c}{when  beta fixed alpha varying} \\
		\hline
		0.5, 0.75 & 1.9392 & 1.1443 & 0.7425 & 3.8789 & 1.4057 & 0.5516 \\
		1.0, 0.75 & 1.2216 & 0.4541 & 0.7425 & 3.8789 & 0.8855 & 0.5516 \\
		1.5, 0.75 & 0.9323 & 0.2645 & 0.7425 & 3.8789 & 0.6758 & 0.5516 \\
		2.5, 0.75 & 0.6632 & 0.1338 & 0.7425 & 3.8789 & 0.4807 & 0.5516 \\
		3.5, 0.75 & 0.5299 & 0.0855 & 0.7425 & 3.8789 & 0.3841 & 0.5516 \\
		\hline
		\multicolumn{7}{c}{when both varying} \\
		\hline
		1, 1  & 1.1287 & 0.2265 & 0.2384 & 3.1071 & 1.0000 & 0.4217 \\
		2, 2  & 0.8723 & 0.0372 & 0.0102 & 2.7895 & 0.8891 & 0.2212 \\
		3, 3  & 0.8484 & 0.0163 & 0.0831 & 2.6907 & 0.8736 & 0.1506 \\
		4, 4  & 0.8509 & 0.0094 & 0.1069 & 1.9643 & 0.8750 & 0.1140 \\
		5, 5  & 0.8586 & 0.0062 & 0.0677 & 0.1072 & 0.8805 & 0.0915 \\
	\hline
	\end{tabular}%
	\label{tab:addlabel}%
\end{table}%

\subsection{Mean Deviation}
The mean deviation (MD) about mean $(\mu^{'}_{1}=\mu)$ is defined by 
\begin{equation}
\begin{split}
MD&=\int_{x}|x-\mu| f(x,\alpha,\beta) dx\\&=\int_{x=0}^{\mu}(\mu-x)f(x,\alpha,\beta) dx+\int_{x=\mu}^{\infty}(x-\mu) f(x,\alpha,\beta) dx
\end{split}
\end{equation}
After simplification, we get
\begin{equation}
\begin{split}
MD&=2\mu F(\mu)-2\mu+2\int_{\mu}^{\infty} f(x,\alpha,\beta) dx\\&=(\mu-1)\left[ \dfrac{4}{\sqrt{\pi}}\gamma\left( \frac{3}{2},\alpha \mu^{2\beta}\right)-2\right] 
\end{split}
\end{equation}
\subsection{Generating Functions}
In distribution theory, the role of generating functions are very usefull to generate the respective moments of the distribution and also these functions are uniquely determine the distribution. The different generating function for PMaD $(\alpha,\beta)$ have been caluculated as follows; 
\begin{itemize}
	\item Moment generating function (mgf) $M_X(t)$ for a random variable $X$  is obatined as
	\begin{equation}
	M_X(t)=E(e^{tx})=\dfrac{2}{\sqrt{\pi}}\sum_{i=0}^{\infty}\frac{1}{j!} \left( \dfrac{t}{\alpha^{2\beta}}\right)^{r}  \Gamma\left( \dfrac{3\beta+r}{2\beta}\right)
	\end{equation}
	\item Characteristics function (chf)  $\phi_X(t)$ for random variable $X$ is obtained by replacing $t$ by $jt$,
		\begin{equation}
	\phi_X(t)=E(e^{jtx})=\dfrac{2}{\sqrt{\pi}}\sum_{i=0}^{\infty}\frac{1}{j!} \left( \dfrac{jt}{\alpha^{2\beta}}\right)^{r}  \Gamma\left( \dfrac{3\beta+r}{2\beta}\right)
	\end{equation}
	where, $j^2=-1$. 
	\item  The kumulants generating function (KGF) is obtained as
		\begin{equation}
	K_X(t)=\ln \left[ \dfrac{2}{\sqrt{\pi}}\sum_{i=0}^{\infty}\frac{1}{j!} \left( \dfrac{t}{\alpha^{2\beta}}\right)^{r}  \Gamma\left( \dfrac{3\beta+r}{2\beta}\right)\right] 
	\end{equation}
\end{itemize}
\subsection{Conditional Moment and MGF}
Let $X$ be a random variable from PMaD$ (\alpha,\beta)$, then conditional moments $E(X^r|X>k)$ and conditional mgf $E(e^{tx}|X>k)$ are evaluated in following expressions;
\begin{equation}
\begin{split}
E(X^r|X>k)&=\dfrac{\int_{x>k}x^r f(x,\alpha,\beta) dx}{\int_{x>k} f(x,\alpha,\beta) dx}\\&=\dfrac{2\left(\frac{1}{\alpha}\right)^{\frac{r}{2\beta}}\gamma\left( \dfrac{3\beta+r}{2\beta},\alpha x^{2\beta}\right)}{\sqrt{\pi}-2\gamma\left( \frac{3}{2},\alpha x^{2\beta}\right)}
\end{split}
\end{equation}
\begin{equation}
\begin{split}
E(e^{tx}|X>k)&=\dfrac{\int_{x>k}e^{tx}f(x,\alpha,\beta) dx}{\int_{x>k} f(x,\alpha,\beta) dx}\\&=\dfrac{2\sum_{i=0}^{\infty}\dfrac{t^i}{i!}\left(\frac{1}{\alpha}\right)^{\frac{r}{2\beta}}\gamma\left( \dfrac{3\beta+r}{2\beta},\alpha x^{2\beta}\right)}{\sqrt{\pi}-2\gamma\left( \frac{3}{2},\alpha x^{2\beta}\right)}
\end{split}
\end{equation}
respectively. 
\subsection{Bonferroni and Lorenz Curves}
In economics to measure the income and poverty level, Bonferroni and Lorenz curves are frequently used. These two have good linkup to each other and has more comprehensive applications in actuarial as well as in demography. It was initially proposed and studied by Bonferroni (1920), matthematically, it is defined as;
\begin{equation}
\zeta(\nu)_{b}=\dfrac{1}{\nu \mu}\int_{0}^{q}xf(x,\alpha,\beta)dx
\end{equation}
\begin{equation}
\zeta(\nu)_{l}=\dfrac{1}{ \mu}\int_{0}^{q}xf(x,\alpha,\beta)dx
\end{equation}
respectively. where $q=F^{-1}(\nu)$ and $\mu=E(X)$. Hence using eqn (2.1), the above two equations are reduces as
\begin{equation}
\zeta(\nu)_{b}=\dfrac{\sqrt{\alpha}~IG\left( \frac{1+2\beta}{2\beta},\alpha q^{2\beta}\right) }{\nu\Gamma\left( \frac{3\beta+1}{2\beta}\right) }
\end{equation}
\begin{equation}
\zeta(\nu)_{l}=\dfrac{\sqrt{\alpha}~IG\left( \frac{1+2\beta}{2\beta},\alpha q^{2\beta}\right) }{\Gamma\left( \frac{3\beta+1}{2\beta}\right) }
\end{equation}
\subsection{Stochastic Ordering}
 A random variable $X$ is said to be stochastically greater $(Y \le_{st} X)$ than $Y$ if $F_X(x)\le F_Y (x)$ for all $x$. In the similar way, $X$ is said to be stochastically greater $(X \le_{st} Y )$ than $Y$ in the
\begin{itemize}
	\item[$\bullet$] hazard rate order $(X  \le_{hr} Y )$ if $h_X(x) \ge h_Y (x)$  $\forall x$.
	\item[$\bullet$] mean residual life order $(X \le_{mrl} Y )$ if $m_X(x) \ge m_Y (x)$ $\forall x$.
	\item[$\bullet$] likelihood ratio order $(X \le_{lr} Y )$ if $\left[\dfrac{f_{X}(x)}{f_{Y}(x)}\right]$ decreases in x.
\end{itemize}
From the above relations, we can veryfied that;
$$(X \le_{lr} Y )\Rightarrow (X \le_{hr} Y )\Downarrow (X \le_{st} Y )\Rightarrow (X \le_{mrl} Y )$$
The PMaD is ordered with respect to the strongest likelihood ratio ordering as shown in the following theorem.\\

\textbf{Theorem:} Let $X\sim PMaD(\alpha_1,\beta_1)$ and $Y\sim PMaD(\alpha_2,\beta_2)$. Then $(X \le_{lr} Y )$ and hence
$(X \le_{hr} Y )$, $(X \le_{mrl} Y )$ and $(X \le_{st} Y )$ for all values of $\alpha_i,\beta_i$; $i=1, 2$.\\

\textbf{Proof:} The likelihood ratio is
 $\left[\dfrac{f_{X}(x)}{f_{Y}(x)}\right]$ i.e.	
\begin{equation*}
\begin{split}
\Phi=\dfrac{f_{X}(x)}{f_{Y}(x)}&=\left( \dfrac{\alpha_1}{\alpha_2}\right)^{\frac{3}{2}} \left( \dfrac{\beta_1}{\beta_2}\right) x^{3(\beta_1-\beta_2)} e^{-(\alpha_1 x^{2\beta_1}+\alpha_2 x^{2\beta_2})}
\end{split}
\end{equation*}
Therefore, 
\begin{equation}
\Phi^{'}=\log \left( \dfrac{f_{X}(x)}{f_{Y}(x)}\right) =\dfrac{1}{x}\left[ 3(\beta_1-\beta_2)-(\alpha_1 x^{2\beta_1}+\alpha_2 x^{2\beta_2})\right] 
\end{equation}
If $\beta_1=\beta_2=\beta (say)$, then $\Phi^{'}<0$, which shows that $(X \le_{lr} Y )$. The remaining ordering behaviour can be proved in same way. Also, if $\alpha_1=\alpha_2=\alpha (say)$ and $\beta_1<\beta_2$ then again $\Phi^{'}<0$, which shows that $(X \le_{lr} Y )$. The remaining ordering can be proved in same way. 
\section{Residual Lifetime}
In survival analysis, the term residual lifetime often used to describe the remaining lifetime associated with any particular system. Here, we derived the expression of residual life and reversed residual life for PMaD. The residual lifetime function is defined by $ R_{t}= P[x -t|x > t], t \ge0$ and the reversed residual life is described as $\bar{R }_{t} = P[t -x|x \le t]$ which denotes the time elapsed from the failure of a component given that its life less or equal to $t$.
\begin{itemize}
	\item \textbf{Residual life time function}\\
	The survival function of the residual lifetime is given by
	\begin{equation}
	S_{R_t}(x)=\dfrac{S(t+x)}{S(t)}= \dfrac{\sqrt{\pi}-2\gamma\left[ \frac{3}{2},\alpha (x+t)^{2\beta}\right] }{\sqrt{\pi}-2\gamma\left( \frac{3}{2},\alpha x^{2\beta}\right) }~~~~~  ;x> t
	\end{equation}
	The corresponding probability density function is
	\begin{equation}
	f_{R_t}(x)= \dfrac{4 \alpha^{\frac{3}{2}}\beta (x+t)^{3\beta-1}e^{-\alpha (x+t)^{2\beta}}}{\sqrt{\pi}-2\gamma\left( \frac{3}{2},\alpha x^{2\beta}\right) }
	\end{equation}
	Thus, hazard function is obtained as
	\begin{equation}
		h_{R_t}(x)=\dfrac{4 \alpha^{\frac{3}{2}}\beta (x+t)^{3\beta-1}e^{-\alpha (x+t)^{2\beta}}}{\sqrt{\pi}-2\gamma\left[ \frac{3}{2},\alpha (x+t)^{2\beta}\right]}
	\end{equation}
	\item  \textbf{Reversed residual lifetime function}\\
	The survival function for MOEIED is given by
	\begin{equation}
	S_{\bar{R}_t}=\dfrac{F(t-x)}{F(t)}= \dfrac{\gamma\left[ \frac{3}{2},\alpha (t-x)^{2\beta}\right] }{\gamma\left( \frac{3}{2},\alpha x^{2\beta}\right) }~~~~~     ;0\le x<t
	\end{equation}
	The associated pdf is evaluated as;
		\begin{equation}
	f_{\bar{R}_t}=\dfrac{2 \alpha^{\frac{3}{2}}\beta (t-x)^{3\beta-1}e^{-\alpha (t-x)^{2\beta}}}{\gamma\left( \frac{3}{2},\alpha x^{2\beta}\right) }
	\end{equation}
	Hence, the hazard function based on reversed residual lifetime is obtained as
	\begin{equation}
h_{\bar{R}_t}=\dfrac{2 \alpha^{\frac{3}{2}}\beta (t-x)^{3\beta-1}e^{-\alpha (t-x)^{2\beta}}}{\gamma\left[ \frac{3}{2},\alpha (t-x)^{2\beta}\right]}
	\end{equation}
\end{itemize}
\section{Entropy Measurements}
In information theory, entropy measurement plays a vital role to study the uncertainty associated with the probability distribution. In this section, we discuss the different measure of change. For more detail about entropy measurement, see Reniyi (1961).
\subsection{Renyi Entropy}
Renyi entropy of a r.v. $x$ is defined as
\begin{equation}
\begin{split}
R_E&=\dfrac{1}{(1-\in)}\ln\left[ \int_{x=0}^{\infty}f_{w}^{\in}(x,\alpha,\beta)dx\right]\\&=\dfrac{1}{(1-\in)}\ln \left[ \int_{x=0}^{\infty}\left\lbrace \dfrac{4}{\sqrt{\pi}} \alpha^{\frac{3}{2}}\beta x^{3\beta-1}e^{-\alpha x^{2\beta}} \right\rbrace^{\in} dx \right] 
\end{split}
\end{equation}
Hence, after solving the internal, we get the following
{\scriptsize \begin{equation}
\begin{split}
R_E&=\dfrac{1}{(1-\in)}\left[\lambda \ln 4-\frac{\lambda}{2}\ln\pi+\lambda\ln\beta-\dfrac{1-\lambda-2\beta}{2\beta}\ln\alpha-\dfrac{3\lambda\beta-\lambda+1}{2\beta}\ln \lambda+\ln \left( \dfrac{3\beta\lambda-\lambda+1}{2\beta}\right)  \right] 
\end{split}
\end{equation}}
\subsection{$\Delta$-Entropy}
The $\beta$-entropy is obtained as follows
\begin{equation}
\Delta_E=\dfrac{1}{\Delta-1}\left[1-\int_{x=0}^{\infty}f^{\Delta}(x,\alpha,\beta)dx \right] 
\end{equation}
Using pdf (1.4) and after simplification the expression for $\beta$-entropy is given by;
\begin{equation}
\Delta_E=\dfrac{1}{\Delta-1}\left[1-\left( \dfrac{4}{\sqrt{\pi}}\right) ^{\Delta}\beta^{\Delta} \left( \dfrac{1}{\alpha}\right)^{\dfrac{1-\Delta-2\beta}{2\beta}} \left( \dfrac{\Gamma\left(\dfrac{3\Delta\beta-\Delta+1}{2\beta} \right) }{\Delta^{\dfrac{3\Delta\beta-\Delta+1}{2\beta}}}\right)  \right] 
\end{equation}
\subsection{Generalized Entropy}
The generalized entropy is obtained by;
\begin{equation}
G_E=\dfrac{\nu_\lambda\mu^{-\lambda}-1}{\lambda(\lambda-1)}~~~~~;\lambda\ne 0,1
\end{equation}
where, $\nu_\lambda=\int_{x=0}^{\infty}x^{\lambda}f_w(x,\alpha,\theta^{})dx$ and $\mu=E(X)$. The value of $\nu_\lambda$ is calculated as

\begin{equation}
\begin{split}
\nu_\lambda&=\dfrac{2}{\sqrt{\pi}}\left( \dfrac{1}{\alpha}\right)^{\frac{\lambda}{2\beta}}\Gamma\left( \dfrac{3\beta+\lambda}{2\beta}\right)  
\end{split}
\end{equation}
After using (4.6) and (2.8), we get
\begin{equation}
G_E=\left( \dfrac{4}{\pi}\right)^{\frac{1-\lambda}{2}} \left[ \dfrac{\Gamma\left( \dfrac{3\beta+\lambda}{2\beta}\right)  \left\lbrace \Gamma\left( \dfrac{3\beta+1}{2\beta}\right)  \right\rbrace  ^{-\lambda}}{\lambda(\lambda-1)}\right] ~~~~~;\lambda\ne0,1
\end{equation}
\section{Parameter Estimation}
Here, we describe maximum likelihood estimation method and Bayes estimation method for estimating the unknown parameters $\alpha, \beta$ of the PMaD. The estimators obtained under these methods are not in nice closed form; thus, numerical approximation techniques are used to get the solution. Further, the performances of these estimators are studied through Monte Carlo simulation.
\subsection{Maximum Likelihood Estimation}
 The most popular and efficient method of classical estimation of the parameter (s) is maximum likelihood estimation. The estimators obtained by this method passes several optimum properties. The maximum likelihood estimation theory required formulation of the likelihood function. Thus, let us suppose that $X_1,X_2,\cdots,X_n$ are the iid random sample of size $n$ taken from PMaD $(\alpha,\beta)$. The likelihood function is written as;
\begin{equation}
\begin{split}
L(\alpha,\theta)&=\prod_{i=1}^{n}\dfrac{4}{\sqrt{\pi}} \alpha^{\frac{3}{2}}\beta x_i^{3\beta-1}e^{-\alpha x_i^{2\beta}}=\dfrac{4^{n}}{\pi^{n/2}} \alpha^{\frac{3n}{2}}\beta^{n} e^{-\alpha\sum_{i=1}^{n} x_i^{2\beta}}\left( \prod_{i=1}^{n}x_i^{3\beta-1}\right) 
\end{split}
\end{equation} 
Log-likelihood function is written as;
\begin{equation}
\ln L(\alpha,\theta)=l=n\ln 4-\frac{n}{2}\ln \pi+\dfrac{3n}{2}\ln \alpha+n\ln\beta
-\alpha\sum_{i=1}^{n}x_i^{2\beta}+(3\beta-1)\sum_{i=1}^{n}\ln x_i
\end{equation}
for MLEs of $\alpha$ and $\beta$, 
$$\dfrac{\partial l}{\partial\alpha}=0~~~\&~~~\dfrac{\partial l}{\partial\beta}=0$$
which yield,
\begin{equation}
\dfrac{3n}{2\alpha}-\sum_{i=1}^{n}x_i^{2\beta}=0
\end{equation}
\begin{equation}
\dfrac{n}{\beta}-2\alpha\sum_{i=1}^{n} x_i^{2\beta}\ln x_i+3\sum_{i=1}^{n}\ln x_i =0
\end{equation}
The MLE's of the parameters are obtained by solving the above two equations simultaneously. Here, we used non-linear maximization techniques to get the solution.
\subsubsection{Uniqueness of MLEs}
The uniqueness of MLEs discussed in previous section can be checked by using  following propositions. \\

\textbf{Proposition 1:} If $\beta$ is fixed, then $\hat{\alpha}$ exist and it is unique. \\

\textbf{Proof:} Let $L_{\alpha}=\dfrac{3n}{2\alpha}-\sum_{i=1}^{n}x_i^{2\beta}$, since $L_{\alpha}$  is continuous and it has been verified that $\lim\limits_{\alpha\rightarrow 0}L_{\alpha}=\infty$ and $\lim\limits_{\alpha\rightarrow \infty}L_{\alpha}=-\sum_{i=1}^{n}x_i^{2\beta}<0$. This implies that $L_{\alpha}$ will have atleast one root in interval $(0,\infty)$ and hence $L_{\alpha}$ is a decreasing function in $\alpha$. Thus, $L_{\alpha}=0$ has a unique solution in $(0,\infty)$. \\

\textbf{Proposition 2:} If $\alpha$ is fixed, then $\hat{\beta}$ exist and it is unique. \\

\textbf{Proof:} Let $L_{\beta}=\dfrac{n}{\beta}-\alpha\sum_{i=1}^{n} x_i^{2\beta}\ln x_i+3\sum_{i=1}^{n}\ln x_i $, since $L_{\beta}$  is continuous and it has been verified that $\lim\limits_{\beta\rightarrow 0}L_{\beta}=\infty$ and $\lim\limits_{\beta\rightarrow \infty}L_{\beta}=-2\sum_{i=1}^{n}\ln {x_i}<0$. This implies in same as above $\hat{\beta}$ exists and it will be unique.
\subsubsection{Fisher Information Matrix}
Here, we derive Fisher information matrix for constructing 100$(1-\Psi)\%$ asymptotic confidence interval for the parameters using large sample theory. The Fisher information matrix can be obtained by using equation (5.2) as
$$ I(\hat{\alpha},\hat{\beta})=-E\begin{pmatrix}
l_{\alpha\alpha} & l_{\alpha\beta}\\
\\
l_{\beta\alpha} & l_{\beta\beta}
\end{pmatrix}_{(\hat{\alpha},\hat{\beta})}  \eqno{(5.2.1)}$$
where, 
$$l_{\alpha\alpha}=-\dfrac{3n}{2\alpha^2},~~l_{\alpha\beta}=-2\sum_{i=1}^{n}x_i^{2\beta}\ln x_i, ~~
l_{\beta\beta}=-\dfrac{n}{\beta^2}-4\alpha\sum_{i=1}^{n}x_i^{2\beta}(\ln x_i)^{2}$$

The above matrix can be inverted and diagonal elements of $I^{-1}(\hat{\alpha},\hat{\beta})$ provide asymptotic variance of $\alpha$ and $\beta$ respectively. Now, two sided $100(1-\Psi)\%$ asymptotic confidence interval for $\alpha$, $\beta$ has been obtained as
$$[\alpha_l,\alpha_u] \in [\hat{\alpha}\mp Z_{1-\frac{\Psi}{2}}\sqrt{var(\hat{\alpha})}]$$
$$[\beta_l,\beta_u] \in [\hat{\beta}\mp Z_{1-\frac{\Psi}{2}}\sqrt{var(\hat{\beta})}]$$
respectively. 
\subsection{Bayes Estimation}
In this subsection, the Bayes estimation procedure for the PMaD has been developed. In this estimation technique, as we all know that the unknown parameter treated as the random variable and this randomness of the parameter quantify in the form of prior distribution. Here, we took two independent gamma prior for both shape and scale parameter. The considered prior is very flexible due to its flexibility of  assuming different shape. Thus, the joint prior $g(\alpha,\beta)$ is given by;

\begin{equation}
g(\alpha,\beta)\propto \alpha^{a-1}\beta^{c-1}e^{-b\alpha-d\beta}~~;~~~\alpha, \beta>0
\end{equation}
  where, a, b, c \& d are the hyperparmaters of the considered priors. Using (5.1) and (5.5), the joint posterior density function $\pi(\alpha,\beta|x)$ is derived as 
\begin{equation}
\begin{split}
\pi(\alpha,\beta|x)&= \dfrac{L(x|\alpha,\beta)g(\alpha,\beta)}{\int_{\alpha}\int_{\beta} L(x|\alpha,\beta)g(\alpha,\beta) d\alpha\,d\beta}\\&= \dfrac{\alpha^{\frac{3n}{2}+a-1}\beta^{n+c-1} e^{-\alpha \left(b+\sum_{i=1}^{n} x_i^{2\beta}\right) }e^{-d\beta}\left( \prod_{i=1}^{n}x_i^{3\beta-1}\right) }{\int_{\alpha}\int_{\beta}\alpha^{\frac{3n}{2}+a-1}\beta^{n+c-1} e^{-\alpha \left(b+\sum_{i=1}^{n} x_i^{2\beta}\right) }e^{-d\beta}\left( \prod_{i=1}^{n}x_i^{3\beta-1}\right) \quad d\alpha\quad d\beta}
\end{split}
\end{equation}
In the Bayesian analysis, the specification of proper loss function plays an important role. Here, we took most frequently used square error loss function (SELF) to obtain the estimates of the parameters. It is defined as;
\begin{equation}
L(\phi,\hat{\phi})\propto \left( \phi-\hat{\phi}\right)^{2} 
\end{equation}
where, $\hat{\phi}$ is estimate of $\phi$. Bayes estimates under SELF is the posterior mean and evaluated by
\begin{equation}
\hat{\phi}_{SELF}=\left[ E(\phi|x)\right]
\end{equation}
provided the expectation exist and finite. Thus, the Bayes estimator based on equation no. (5.6) under SELF are given by
\begin{equation}
\hat{\alpha}_{bs}=E_{\alpha,\beta|x}(\alpha|\beta,x)=\eta \int_{\alpha}\int_{\beta}\alpha^{\frac{3}{2}+a}\beta^{n+c-1} e^{-\alpha \left(b+\sum_{i=1}^{n} x_i^{2\beta}\right) }e^{-d\beta}\left( \prod_{i=1}^{n}x_i^{3\beta-1}\right)  ~ d\alpha~ d\beta
\end{equation} 
and
\begin{equation}
\hat{\beta}_{bs}=E_{\alpha,\beta|x}(\beta|\alpha,x)=\eta \int_{\alpha}\int_{\beta}\alpha^{\frac{1}{2}+a}\beta^{n+c} e^{-\alpha \left(b+\sum_{i=1}^{n} x_i^{2\beta}\right) }e^{-d\beta}\left( \prod_{i=1}^{n}x_i^{3\beta-1}\right)  ~ d\alpha~ d\beta
\end{equation} 
where, $\eta=\int_{\alpha}\int_{\beta}\alpha^{\frac{3n}{2}+a-1}\beta^{n+c-1} e^{-\alpha \left(b+\sum_{i=1}^{n} x_i^{2\beta}\right) }e^{-d\beta}\left( \prod_{i=1}^{n}x_i^{3\beta-1}\right) ~d\alpha~ d\beta$ \\

From equation number  (5.9), (5.10) it is easy to observed that the posterior expectations are appearing in the form of the ratio of two integrals. Thus, the analytical solution of these expetations are not presumable. Therefore, any numerical approximation techniques may be implemented to secure the solutions. Here, we used one of the most popular  and quite effective approximation technique suggested by Lindley (1980). The detailed description can be seen in below;
 \begin{eqnarray}\label{eq20}
 (\hat{\alpha},\hat{\beta})_{Bayes}&=&\dfrac{\int_{\alpha}\int_{\beta}u(\alpha,\beta)e^{\rho(\alpha,\beta)+l}\quad d\alpha d\beta }{\int_{\alpha}\int_{\beta}e^{\rho(\alpha,\beta)+l}\quad d\alpha d\beta}\\&=&(\hat{\alpha},\hat{\beta})_{ml}+\dfrac{1}{2}[(u_{\alpha\alpha}+ 2 u_\alpha\rho_\alpha)\tau_{\alpha\alpha}+(u_{\alpha\beta}+2u_\alpha\rho_\beta)\tau_{\alpha\beta}+(u_{\beta\alpha}+ 2u_\beta\rho_\alpha)\tau_{\beta\alpha}\nonumber\\
 &+&(u_{\beta\beta}+2u_\beta\rho_\beta)\tau_{\beta\beta}]+\dfrac{\alpha}{\beta}[(u_\alpha\tau_{\alpha\alpha}+u_\beta\tau_{\alpha\beta})(l_{111}\tau_{\alpha\alpha}+2l_{21}\tau_{\alpha\beta}+l_{12}\tau_{\beta\beta})\nonumber\\
 &+&(u_\alpha\tau_{\beta\alpha}+u_\beta\tau_{\beta\beta})(l_{21}\tau_{\alpha\alpha}+2 l_{12}\tau_{\beta\alpha}+l_{222}\tau_{\beta\beta})]
 \end{eqnarray}
 where, $u(\alpha,\beta)=(\alpha,\beta)$,  $\rho(\alpha,\beta)=\ln g(\alpha,\beta)$ and $l=\ln L(\alpha,\beta|\underbar x)$,
 \begin{eqnarray*}
 	l_{ab}=\dfrac{\partial^{3}l}{\partial\alpha^{a}\partial\beta^{b}},\quad a,b=0,1,2,3\quad a+b=3,\quad
 	\rho_\alpha=\dfrac{\partial \rho}{\partial\alpha},\quad\rho_\beta=\dfrac{\partial \rho}{\partial\beta}\\ u_\alpha=\dfrac{\partial u}{\partial\alpha},\quad u_\beta=\dfrac{\partial u}{\partial \beta},\quad u_{\alpha\alpha}=\dfrac{\partial^2 u}{\partial\alpha^2},\quad u_{\beta\beta}=\dfrac{\partial^2 u}{\partial\beta^2},\quad u_{\alpha\beta}=\dfrac{\partial^2 u}{\partial\alpha\partial\beta},
 \end{eqnarray*}
$$\tau_{\alpha\alpha}=\dfrac{1}{l_{20}},\,\tau_{\alpha\beta}=\dfrac{1}{l_{11}}=\tau_{\beta\alpha},\, \tau_{\beta\beta}=\dfrac{1}{l_{02}}$$

%
 since, $u(\alpha,\beta$ is the function of $\alpha,\beta$ both. Therefore,
 \begin{itemize}
 	\item If $u(\alpha,\beta)=\alpha$ in (5.12) then;
  \begin{align*}
 u_\alpha&=1,\quad u_\beta=0, \quad u_{\alpha\alpha}=u_{\beta\beta}=0,\quad u_{\alpha\beta}=u_{\beta\alpha}=0
 \end{align*}
 \item If $u(\alpha,\beta)=\beta$ in (5.12) then;
 \begin{align*}
 u_\beta&=1,\quad u_\alpha=0, \quad u_{\alpha\alpha}=u_{\beta\beta}=0,\quad u_{\alpha\beta}=u_{\beta\alpha}=0\\
 \end{align*}
\end{itemize}
and the rest derivatives based on likelihood function are as obtained as; 
$$l_{30}=\dfrac{3n}{\alpha^3},~~l_{11}=-2\sum_{i=1}^{n}x_i^{2\beta}\ln x_i, ~~
l_{03}=\dfrac{2n}{\beta^3}-8\alpha\sum_{i=1}^{n}x_i^{2\beta}(\ln x_i)^{3}$$
$$l_{12}=-4\sum_{i=1}^{n}x_i^{2\beta}(\ln x_i)^{2}=l_{21}$$
Using these derivatives the Bayes estimates of $(\alpha,\beta)$ are obtained by following expressions
\begin{equation}
\begin{split}
\hat{\alpha}_{bl}=&\hat{\alpha}_{ml}+\dfrac{1}{2}[(2u_\alpha\rho_\alpha)\tau_{\alpha\alpha}+(2u_\alpha\rho_\beta)\tau_{\alpha\beta}]+\dfrac{1}{2}[(u_\alpha\tau_{\alpha\alpha})(l_{30}\tau_{\alpha\alpha}+2l_{21}\tau_{\alpha\beta}+l_{12}\tau_{\beta\beta})\\&+(u_\alpha\tau_{\beta\alpha})(l_{21}\tau_{\alpha\alpha}+2l_{12}\tau_{\beta\alpha}+l_{03}\tau_{\beta\beta})]
\end{split}
\end{equation}
\begin{equation}
\begin{split}
\hat{\beta}_{bl}&=\hat{\beta}_{ml}+\dfrac{1}{2}[(2u_\beta\rho_\alpha)\tau_{\beta\alpha}+(2u_\beta\rho_\beta)\tau_{\beta\beta}]+\dfrac{1}{2}[(u_\beta\tau_{\alpha\beta})(l_{30}\tau_{\alpha\alpha}+2l_{21}\tau_{\alpha\beta}+l_{12}\tau_{\beta\beta})\\&+(u_\beta\tau_{\beta\beta})(l_{21}\tau_{\alpha\alpha}+2l_{12}\tau_{\beta\alpha}+l_{03}\tau_{\beta\beta})]
\end{split}
\end{equation}

\section{Simulation Study}
In this section, Monte Carlo simulation study has been performed to assess the performance of the obtained estimators  in terms of their mean square error (MSEs). The maximum likelihood estimates of the parameters are evaluated by using $nlm () $ function, and MLEs of reliability characteristics are obtained by using invariance properties. The Bayes estimates of the parameter are evaluated by Lindley's approximation technique. The  hyper-parameters values are chosen in such a way that the prior mean is equal to the true value, and prior variance is taken as very small, say 0.5. All the computations are done by $R 3.4.1$ software. At first, we generated 5000 random samples from PMaD $(\alpha,\beta)$ using Newton-Raphson algorithm for different variation of sample sizes as $n=10$ (small), $n=20, 30$ (moderate), $n=50$ (large) for fixed $(\alpha=0.75, \beta=0.75)$ and secondly for different variation of $(\alpha, \beta)$ when sample size is fixed say $(n=20)$  respectively. Average estimates and mean square error (MSE) of the parameters and reliability characteristics are calculated for the above mentioned choices, and the corresponding results are reported in Table 2. The asymptotic confidence interval (ACI) and asymptotic confidence length (ACL) are also obtained and presented in Table 3.  From this extensive simulation study, it has been observed that the precision of MLEs and Bayes estimator are increasing when the sample size is increasing while average ACL is decresing. Further, the Bayes estimators are more precise as compared ML estimators for all considered cases.
\begin{table}[h]
	\centering
	\caption{Average estimates and mean square errors (in each second row) of the parameters and reliability characteristics based on simulated data}
	\begin{tabular}{ccccccccc}\\
		\hline
		n     & $\alpha, \beta$ & $\alpha_{ml}$& $\beta_{ml}$ & $MTTF_{ml}$ &$R (t)_{ml}$& $H(t)_{ml}$ & $\alpha_{bl}$ & $\beta_{bl}$ \\
		\hline
		\multirow{2}[0]{*}{10} & \multirow{8}[0]{*}{0.75,0.75} & 0.5070 & 1.1598 & 1.5119 & 0.9691 & 0.1663 & 0.5063 & 1.1028 \\
		&       & 0.0631 & 0.2588 & 0.0164 & 0.0049 & 0.0947 & 0.0631 & 0.2027 \\
		\multirow{2}[0]{*}{20} &       & 0.6560 & 0.8848 & 1.4922 & 0.9343 & 0.2965 & 0.6521 & 0.8647 \\
		&       & 0.0098 & 0.0326 & 0.0093 & 0.0014 & 0.0703 & 0.0105 & 0.0263 \\
		\multirow{2}[0]{*}{30} &       & 0.7096 & 0.8064 & 1.4883 & 0.9163 & 0.3504 & 0.7058 & 0.7951 \\
		&       & 0.0022 & 0.0103 & 0.0071 & 0.0004 & 0.0010 & 0.0025 & 0.0087 \\
		\multirow{2}[0]{*}{50} &       & 0.7542 & 0.7453 & 1.4869 & 0.8988 & 0.3968 & 0.7514 & 0.7397 \\
		&       & 0.0003 & 0.0031 & 0.0046 & 0.0001 & 0.0003 & 0.0003 & 0.0031 \\
		\hline
		\multicolumn{9}{c}{for fixed n and different alpha, beta} \\
		\hline
		\multirow{8}[0]{*}{20} & \multirow{2}[0]{*}{0.5,0.75} & 0.6603 & 0.6832 & 1.7380 & 0.9044 & 0.3400 & 0.6574 & 0.6716 \\
		&       & 0.0261 & 0.0125 & 0.0585 & 0.0017 & 0.0099 & 0.0252 & 0.0117 \\
		& \multirow{2}[0]{*}{0.5, 1.5} & 0.7290 & 0.3033 & 4.6222 & 0.7871 & 0.3556 & 0.7258 & 0.3229 \\
		&       & 0.0528 & 1.4330 & 11.9171 & 0.0402 & 0.1139 & 0.0513 & 1.3866 \\
		& \multirow{2}[0]{*}{1.5, 0.5} & 0.5090 & 2.9297 & 1.1531 & 0.9983 & 0.0207 & 0.5517 & 2.8634 \\
		&       & 0.9907 & 6.6465 & 0.0242 & 0.1274 & 26.0695 & 0.9087 & 6.3006 \\
		& \multirow{2}[0]{*}{2.5,2.5} & 1.0448 & 0.5958 & 1.4084 & 0.7953 & 0.6393 & 1.2825 & 0.6727 \\
		&       & 2.1402 & 3.6573 & 0.3860 & 0.0373 & 0.3553 & 1.5058 & 3.3715 \\
		\hline
	\end{tabular}%
	\label{tab:addlabel}%
\end{table}%
\begin{table}[htbp]
	\centering
	\caption{Interval estimates and asymptotic confidence length (ACL) of the parameters}
	\begin{tabular}{cccccccc}\\
		\hline
		n     & $\alpha, \beta$ & $\alpha_L$ & $\alpha_U$ & $ACL_\alpha$ & $\beta_L$& $\beta_U$ & $ACl_\beta$ \\
		\hline
		10    & 0.75,0.75 & 0.0874 & 0.9266 & 0.8393 & 0.5711 & 1.7485 & 1.1775 \\
		20    & 0.75,0.75 & 0.3209 & 0.9911 & 0.6703 & 0.5525 & 1.2171 & 0.6646 \\
		30    & 0.75,0.75 & 0.4263 & 0.9928 & 0.5665 & 0.5555 & 1.0574 & 0.5019 \\
		50    & 0.75,0.75 & 0.5290 & 0.9794 & 0.4505 & 0.5631 & 0.9275 & 0.3644 \\
		\hline
		\multicolumn{8}{c}{for fixed n and different $\alpha, \beta$} \\
		\hline
		\multirow{4}[0]{*}{20} & 0.5, 0.75 & 0.3255 & 0.9951 & 0.6696 & 0.4142 & 0.9523 & 0.5381 \\
		& 0.5, 1.5 & 0.3794 & 1.0785 & 0.6991 & 0.4819 & 1.7425 & 1.2429 \\
		& 1.5, 0.5 & 0.4206 & 1.7812 & 0.76058 & 0.2260 & 1.8334 & 1.3807 \\
		& 2.5, 2.5 & 0.5804 & 2.9509 & 0.9788 & 0.54133 & 2.7783 & 1.1365 \\
	\hline
	\end{tabular}%
	\label{tab:addlabel}%
\end{table}%
\section{Real Data Illustration}
This section,  demonstrate the practical applicability of the proposed model in real life scenario especially for the survival/relibaility data taken from diffierent sources. The proposed distribution is compared with Maxwell distribution (MaD) and its different generalizations, such as length biased maxwell distribution (LBMaD), area biased maxwell distribution (ABMaD), extended Maxwell distribution (EMaD) and generalized Maxwell distribution (EMaD). For these models the estimates of the parameter (s) are obtained by method of maximum likelihood and the compatibility of PMaD has been discussed using model selection tools such as log-likelihood (-log L), Akaike information criterion (AIC), corrected Akaike information criterion (AICC), Bayesian information criterion (BIC) and Kolmogorov Smirnov (K-S) test. In general, the smaller values of these statistics indicate, the better fit to the data. The point estimates of the parameters and reliability function and hazard function for each data set are reported in Table 6. The interval estimate of the parameter and corresponding asimptotic confidence length  are also evaluated and presented in Table 7.

\textbf{Data Set-I (Bladder Cancer Data):} This data set represents the remission times (in months) of a 128 bladder cancer patients, and it was initially used by Lee and Wang (2003). The same data set is used to show the superiority of extended maxwell distribution by  Sharma et al (2017). \\

\textbf{Data Set-II : Item Failure Data}\\
 This dataset is taken from Murthy et al. (2004). It shows 50 items put into use at t = 0 and failure items are recorded in weeks.\\

\textbf{Data Set-III:  } The data set was initially considered by Chhikara and Folks (1977). It represent the 46 repair times (in hours) for an airborne communication transceiver.\\

\textbf{Data Set-IV: Flood data}\\
The data are the exceedances of flood peaks (in m3/s) of the Wheaton River near Carcross
in Yukon Territory, Canada. The data consist of 72 exceedances for the years 1958–1984, rounded to one decimal place. This data was analyzed by Choulakian and Stephens (2011).
\begin{table}[htbp]
	\centering
	\caption{Goodness of fit values for different model}
	\begin{tabular}{cccccccc}\\
		\hline
		\multicolumn{8}{c}{Bladder cancer data N=128} \\
		\hline
	Model & $ \hat{\alpha} $ &  $ \hat{\beta} $ & -logL  & AIC   & AICC  & BIC   & K-S \\
	\hline
		\textbf{PMaD} & \textbf{0.7978} & \textbf{0.1637} & \textbf{366.3820} & \textbf{736.7639} & \textbf{732.8599} & \textbf{742.4680} & \textbf{0.3675} \\
		MaD   & 0.0076 &--    & 1014.4440 & 2030.8870 & 2028.9190 & 2033.7400 & 0.4144 \\
		LBMaD & 98.6386 & --    & 669.3668 & 1340.7340 & 1338.7650 & 1343.5860 & 0.4906 \\
		ABMaD & 78.9109 & --   & 767.8122 & 1537.6240 & 1535.6560 & 1540.4770 & 0.5608 \\
		ExMaD & 0.8447 & 1.4431 & 412.1232 & 828.2464 & 824.3424 & 833.9504 & 0.8265 \\
		GMaD  & 0.7484 & 527.2314 & 426.6019 & 857.2037 & 853.2997 & 862.9078 & 0.7086 \\
		\hline
		\multicolumn{8}{c}{Item failure data N=50} \\
		\hline
	Model & $ \hat{\alpha} $ &  $ \hat{\beta} $ & -logL  & AIC   & AICC  & BIC   & K-S \\
	\hline
		\textbf{PMaD} & \textbf{0.8339} & \textbf{0.1820} & \textbf{135.8204} & \textbf{275.6407} & \textbf{271.8961} & \textbf{279.4648} & \textbf{0.2625} \\
		MaD   & 0.0104 & --   & 367.8528 & 737.7056 & 735.7890 & 739.6177 & 0.4268 \\
		LBMaD & 72.1146 & --   & 315.1624 & 632.3248 & 630.4081 & 634.2368 & 0.5112 \\
		ABMaD & 57.6917 & --    & 374.1247 & 750.2494 & 748.3328 & 752.1615 & 0.5825 \\
		ExMaD & 0.6186 & 1.0139 & 151.2998 & 306.5996 & 302.8550 & 310.4237 & 0.7327 \\
		GMaD  & 0.5400 & 534.1569 & 151.2643 & 306.5287 & 302.7840 & 310.3527 & 0.3920 \\
		\hline
		\multicolumn{8}{c}{Airborne communication transceiver N=46} \\
		\hline
	Model & $ \hat{\alpha} $ &  $ \hat{\beta} $ & -\ logL   & AIC   & AICC  & BIC   & K-S \\
	\hline
		\textbf{PMaD} & \textbf{0.8735} & \textbf{0.2709} & \textbf{101.9125} & \textbf{207.8249} & \textbf{204.1040} & \textbf{211.4822} & \textbf{0.2136} \\
		MaD   & 0.0406 & --   & 245.1383 & 492.2766 & 490.3675 & 494.1052 & 0.5027 \\
		LBMaD & 18.4603 & --   & 237.4945 & 476.9890 & 475.0799 & 478.8176 & 0.5771 \\
		ABMaD & 14.7683 & --   & 284.7017 & 571.4034 & 569.4943 & 573.2320 & 0.6324 \\
		ExMaD & 0.7290 & 0.8672 & 103.3052 & 210.6104 & 206.8895 & 214.2677 & 0.2989 \\
		GMaD  & 0.6015 & 122.7666 & 110.8521 & 225.7042 & 221.9833 & 229.3615 & 0.4392 \\
		\hline
		\multicolumn{8}{c}{River data N=72} \\
		\hline
Model & $ \hat{\alpha} $ &  $ \hat{\beta} $ & - logL  & AIC   & AICC  & BIC   & K-S \\
\hline
		\textbf{PMaD} & 0.805185 & 0.1504145 & 212.8942 & 429.7884 & 425.9623 & 434.3418 & 0.2760 \\
		MaD   & 0.005032 & --    & 610.9235 & 1223.847 & 1221.904 & 1226.124 & 0.3821 \\
		LBMaD & 149.0315 & --    & 426.3076 & 854.6153 & 852.6724 & 856.8919 & 0.4113 \\
		ABMaD & 119.2252 & --   & 493.3271 & 988.6543 & 986.7114 & 990.9309 & 0.4529 \\
		ExMaD & 0.697471 & 1.306933 & 251.9244 & 507.8487 & 504.0226 & 512.4021 & 0.7487 \\
		GMaD  & 0.648149 & 919.7356 & 251.2767 & 506.5534 & 502.7273 & 511.1068 & 0.4998 \\
		\hline
	\end{tabular}%
	\label{tab:addlabel}%
\end{table}%
\begin{table}[htbp]
	\centering
	\caption{Summary of the data sets}
	\begin{tabular}{ccccccccc}\\
		\hline
		Data  & Min   & Q1    & Q2    & Mean  & Q3    & Max   & Kurtosis & Skewness \\
		\hline
		I     & 0.080 & 3.348 & 6.395 & 9.366 & 11.838 & 79.050 & 18.483 & 3.287 \\
		II    & 0.013 & 1.390 & 5.320 & 7.821 & 10.043 & 48.105 & 9.408 & 2.306 \\
		III   & 0.200 & 0.800 & 1.750 & 3.607 & 4.375 & 24.500 & 11.803 & 2.888 \\
		IV    & 0.100 & 2.125 & 9.500 & 12.204 & 20.125 & 64.000 & 5.890 & 1.473 \\
		\hline
	\end{tabular}%
	\label{tab:addlabel}%
\end{table}%
\begin{table}[htbp]
	\centering
	\caption{Real data estimates}
	\begin{tabular}{cccccccc}\\
		\hline
		Data  & $ \alpha_{ml}$& $ \beta_{ml}$ & $ MTTF_{ml} $ & $ R(t)_{ml} $ & $ H(t)_{ml} $&$ \alpha_{bl}$ &$ \beta_{bl}$ \\
	\hline
		I     & 0.7978 & 0.1637 & 28.2109 & 0.7019 & 0.2827 & 0.7962 & 0.1639 \\
		II    & 0.8339 & 0.1820 & 15.3594 & 0.6953 & 0.3224 & 0.8292 & 0.1821 \\
		III   & 0.8735 & 0.2709 & 4.0773 & 0.7212 & 0.4326 & 0.8675 & 0.2703 \\
		IV    & 0.8052 & 0.1504 & 42.8622 & 0.6923 & 0.2696 & 0.8023 & 0.1506 \\
		\hline
	\end{tabular}%
	\label{tab:addlabel}%
\end{table}%
\begin{table}[htbp]
	\centering
	\caption{Interval estimates based on real data}
	\begin{tabular}{ccccccc}\\
		\hline
		Data  & $ \alpha_L $ & $ \alpha_U $ & $ ACL_\alpha $ & $ \beta_L $ & $ \beta_U $ & $ ACL_\beta $ \\
		\hline
		I     & 0.6545 & 0.9411 & 0.2866 & 0.1373 & 0.1902 & 0.0529 \\
		II    & 0.5962 & 1.0717 & 0.4754 & 0.1376 & 0.2263 & 0.0888 \\
		III   & 0.6202 & 1.1269 & 0.5067 & 0.2081 & 0.3337 & 0.1256 \\
		IV    & 0.6126 & 0.9978 & 0.3852 & 0.1186 & 0.1822 & 0.0636 \\
		\hline
	\end{tabular}%
	\label{tab:addlabel}%
\end{table}%

\begin{figure}
	\centering
	\includegraphics[width=6.5in,height=3in]{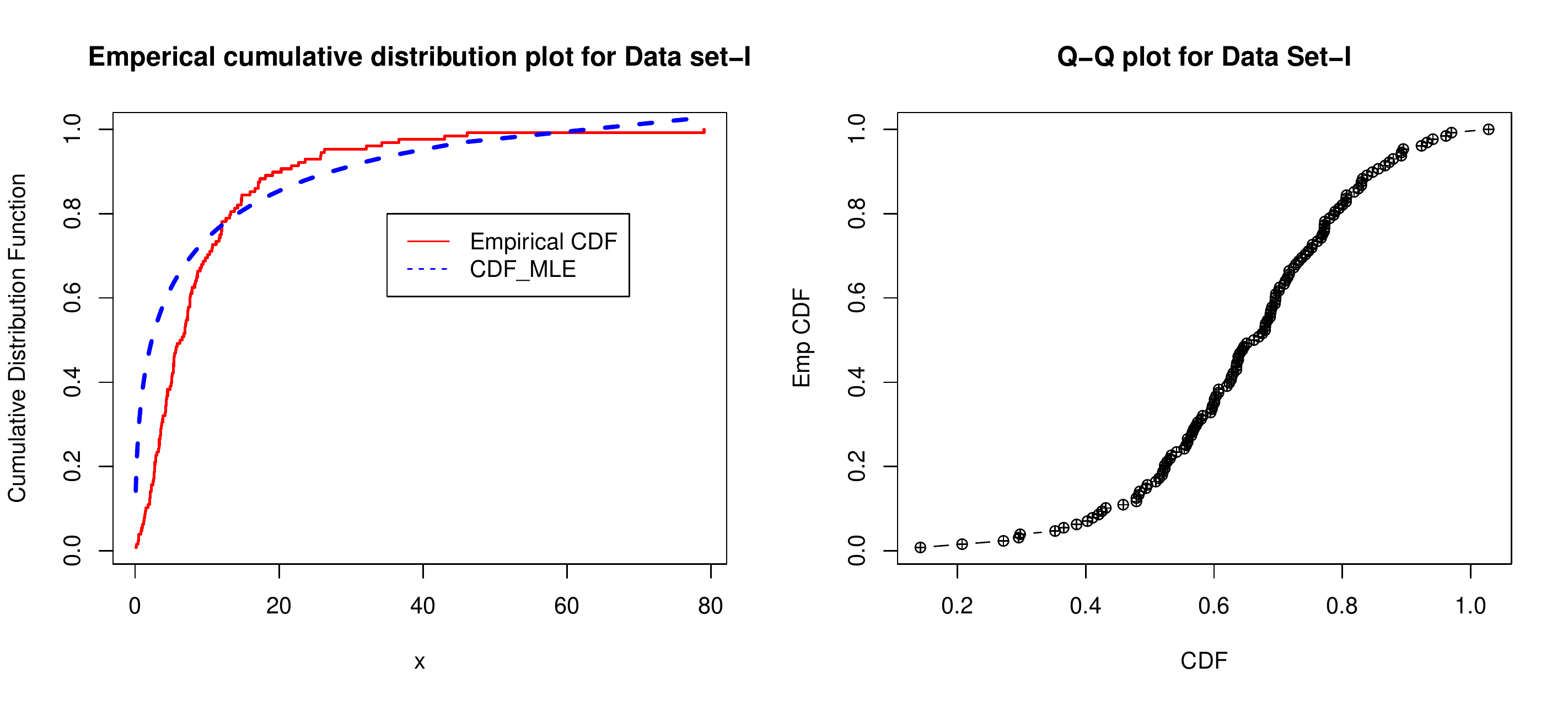}
	\caption{Empirical cumulative distribution function and QQ plot for the data set-I}
\end{figure}
\begin{figure}
	\centering
	\includegraphics[width=6.5in,height=3in]{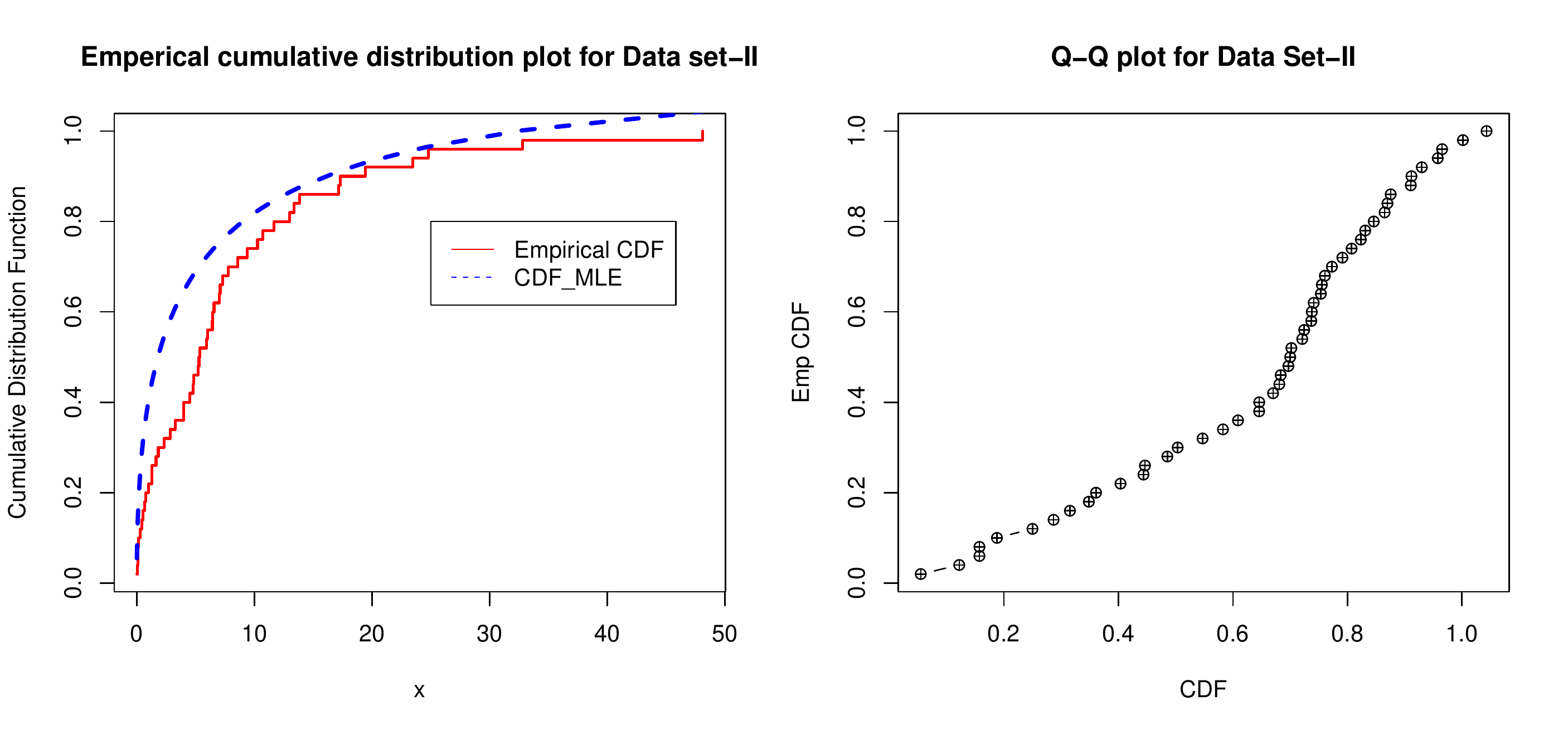}
	\caption{Empirical cumulative distribution function and QQ plot for the data set-II}
\end{figure}
\begin{figure}
	\centering
	\includegraphics[width=6.5in,height=3in]{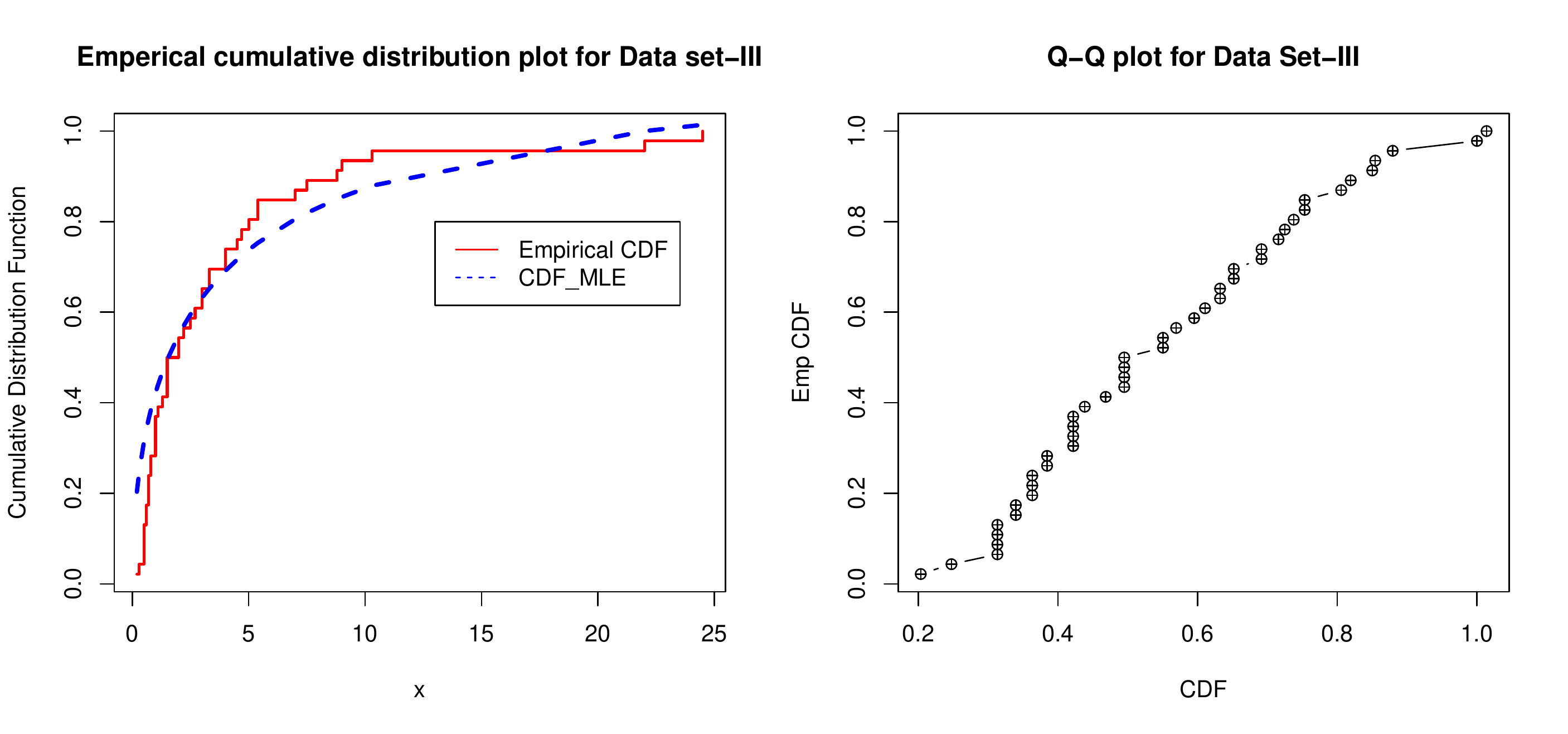}
	\caption{Empirical cumulative distribution function and QQ plot for the data set-III}
\end{figure}
\begin{figure}
	\centering
	\includegraphics[width=6.5in,height=3in]{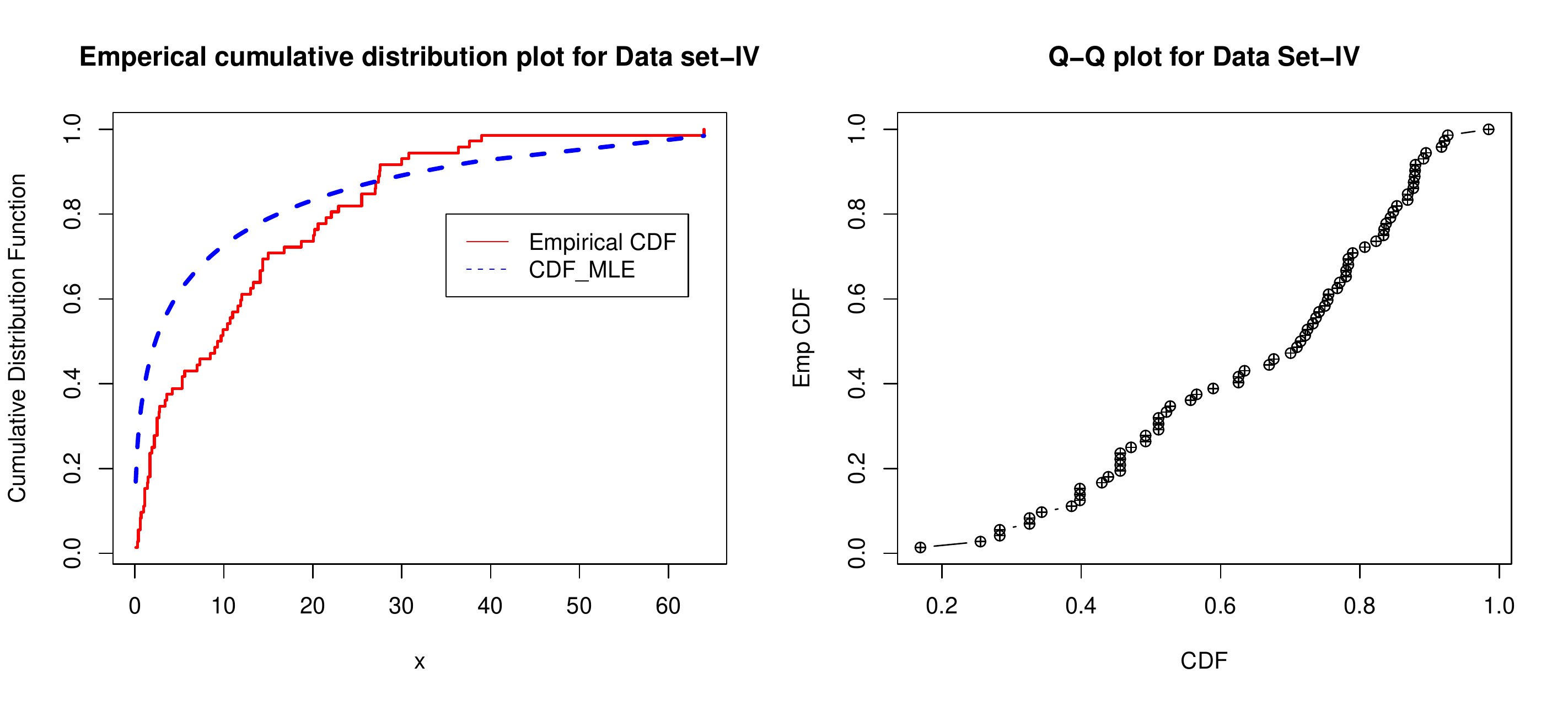}
	\caption{Empirical cumulative distribution function and QQ plot for the data set-IV}
\end{figure}
From the tabulated value of different model selection tools, -LogL, AIC, AICC,  BIC,  and K-S values it has been noticed that PMaD has least -LogL, AIC, AICC,  BIC,  and K-S. The empirical cumulative distribution function and Q-Q plots are also given in Figure 5, 6 \& 7 respectively. Therefore, PMaD can be recommended as a good alternative to the existing family of Maxwell distribution. Summary of the considered data sets is given in Table 2 and seen that skewness is positive for all data sets which indicates that it has positive skewness which appropriately suited to the proposed model.
\section{Conclusion}
This article proposed power Maxwell distribution (PMaD) as an extension of Maxwell distribution and studied its different mathematical and statistical properties, such as reliability characteristics, moments, median, mode, mean deviation, generating functions, stochastic ordering, residual functions, entropy etc. We also study the skewness and kurtosis of the
PMaD and found that it is capable of modeling the positively skewed as well as symmetric data sets. The unknown parameters of the PMaD are estimated by maximum likelihood estimation and Bayes estimation method. The MLEs of the reliability function and hazard function are also obtained by using invariance property. The 95\% asymptotic confidence interval for the parameter are constructed using Fisher information matrix. The MLEs and Bayes estimators are compared through the Monte Carlo simulation and observed that Bayes estimators are more precise under informative prior. Finally, medical/reliability data have been used to show practical utility of the power Maxwell distribution, and it is observed that PMaD provides the better fit as compared to other Maxwell family of distributions. Thus, it can be recommended as an alternative model for the non-monotone failure rate model. 

\newpage

\end{document}